\documentstyle[floats,twoside,eqsecnum,fleqn,aps]{revtex}
\newif\ifepsf \global\epsffalse

\ifepsf
\input{epsf.sty}
\else
\newlength{\epsfxsize}
\newcommand{\epsffile}[1]{%
	\framebox[\epsfxsize]{%
	\centerline{{\tt #1}}}%
}
\fi

\def\mathrm#1{{\rm #1}}
\def\d{\,\mathrm{d}}
\def\e{\mathop{\mathrm{e}}\nolimits}
\def\vec#1{\bbox{#1}}
\def\Tr{\mathop{\mathrm{Tr}}\nolimits}
\mathchardef\varDelta="0101
\mathchardef\varLambda="0103
\mathchardef\varSigma="0106
\mathchardef\varOmega="010A

\begin{document}
\title{
Classical Bifurcation and Enhancement of Quantum Shells
\hbox{\it --- Systematic Analysis of Reflection-Asymmetric Deformed
Oscillator ---}
}
\author{
Ken-ichiro {\sc Arita}%
\thanks{Present address: {\it Yukawa Institute for Theoretical Physics,
Kyoto University, Kyoto 606-01, Japan.}}
and Kenichi {\sc Matsuyanagi}}
\address{
Department of Physics, Kyoto University, Kyoto 606-01, Japan}
\date{April, 1995}
\maketitle
\bigskip

\begin{abstract}
Correspondence between classical periodic orbits and quantum shell
structure is investigated for a reflection-asymmetric deformed
oscillator model as a function of quadrupole and octupole deformation
parameters.  Periodic orbit theory reveals several aspects of quantum
level structure for this non-integrable system.  Good
classical-quantum correspondence is obtained in the Fourier transform
of the quantum level density, and importance of periodic orbit
bifurcation is demonstrated.  Systematic survey of the local minima of
shell energies in the two-dimensional deformation parameter space
shows that prominent shell structures do emerge at finite values of
the octupole parameter.  Correspondences between the regions
exhibiting strong shell effects and the classical bifurcation lines
are investigated, and significance of these bifurcations is indicated.
\end{abstract}

\tableofcontents

\section{\label{SEC:INTRO}Introduction}

Shell structure is one of the important aspects of finite quantum
many-body systems.  In the single-particle level density, one may
generally find some regular patterns like shells consisting of dense
and thin regions.  This pattern changes with deformation, and the
system favors the shape which makes the level density at the Fermi
surface lower.  Predictions of the superdeformed (extremely large
quadrupole deformations with axis ratio about 2:1) and the
hyperdeformed (the axis ratio about 3:1) nuclear states, which are hot
current topics of high-spin nuclear structure physics, had been based
on the above kind of consideration.  Strong shell effects which appear
at ellipsoidal shapes with the axis ratios about 2:1 and 3:1 play
essential roles in stabilizing such exotic shapes.
Reflection-asymmetric degrees of freedom are also one of the most
exciting subjects in the current high-spin physics.  Superdeformed
potentials possess remarkable single-particle level structures where
levels with different parities approximately (exactly in the harmonic
oscillator limit) degenerate in the same major shell, which may bring
about strong octupole correlations \cite{AFN:90,NO:84,MNA:93}.  Recent
remarkable development of large $\gamma$-ray detector arrays
encourages us to find such exotic shapes like reflection-asymmetric
superdeformations, and to investigate the mechanism of producing them.
Recently micro-cluster physics have also attracted much attentions,
and many nuclear physicists have been contributing to this new field.
Shell structures and deformations of clusters are very interesting
subjects --- their shapes can actually be seen with an electron
microscope --- , and one can apply almost the same theoretical
framework to both nuclei and micro-clusters \cite{NHM:90,FP:93,HN:94}.

A clear understanding of the origin of shell structure may be obtained
by the use of semiclassical theory.  Correspondences between quantum
spectra and classical dynamical properties of Hamiltonian systems have
been extensively investigated for two limiting cases, namely, for
integrable and strongly chaotic situations.  Most physical systems are
situated in the midst of these limits, however, and belong to what we
call `mixed' systems.  The semiclassical theory for mixed systems is
difficult and only few aspects have been clarified up to now.  This
difficulty is associated with the periodic-orbit bifurcations
(characteristic to the mixed systems) where the stationary phase
approximation (SPA) and the conventional trace formula for
representing the quantum spectrum in terms of classical periodic
orbits breaks down.  Fortunately, however, an approach in the inverse
direction sometimes works and one can extract the periodic orbit
information from the quantum spectrum by means of the Fourier
transformations.  This approach is very useful to understand the shell
structure of the quantum spectrum.  We take this approach and clarify
some aspects of a mixed system, directing our attention to the
influence of the bifurcations of short periodic orbits on the gross
structure of the quantum spectrum.

In this paper, we investigate the classical-quantum correspondence for
an axially-symmetric deformed oscillator model with
reflection-asymmetric terms.  This is a non-integrable model and
chaotic behavior gradually emerges in the dynamics as the octupole
deformation becomes large.  The role of periodic orbit bifurcations
will be emphasized.  In section \ref{SEC:FORM}, basic elements of the
semiclassical theory relevant to our analysis are briefly reviewed.
Special attention will be paid on the classical bifurcation phenomena
and their effects on the quantum spectra.  In section \ref{SEC:MODEL}
our model is introduced and several aspects of it are summarized.  In
sections \ref{SEC:ESH}$\sim$\ref{SEC:SD}, we will present numerical
results of the semiclassical analysis and discuss their implications.
It will be shown that prominent shell structures emerge for finite
octupole deformations superposed on the prolate shapes.  Origins of
such new shell structures will be clarified.  Section
\ref{SEC:SUMMARY} is devoted to summary and conclusion.

\section{\label{SEC:FORM}Basic formulae}

\subsection{\label{SEC:CL}Classical Hamiltonian system}

Let us consider a Hamiltonian system with $f$ degrees of freedom.
The equation of motion (EOM) for a $2f$-dimensional phase
space vector $Z=(\,\vec{p},\vec{q})$ is expressed as
\begin{equation}
	\frac{\d}{\d t}\,Z
	=\varLambda\,\frac{\partial H}{\partial Z},
	\qquad \varLambda=\left(
	\begin{array}{c@{\quad}c}
		O & -I \\ I & O
	\end{array}\right),
\end{equation}
where $O$ and $I$ denote the $f$-dimensional zero and identical
matrices, respectively.  Now consider a bundle of trajectories around
a certain solution $Z_\alpha(t)$ and write them as
$Z(t)=Z_\alpha(t)+\delta Z(t)$.  Then the EOM for $\delta Z(t)$ is
given by
\begin{equation}
	\frac{\d}{\d t}\,\delta Z
	=\varLambda\,{\cal H}(Z_\alpha(t))\,\delta Z,
	\qquad {\cal H}(Z)_{ij}
	=\frac{\partial^2 H(Z)}{\partial Z_i\partial Z_j}
\end{equation}
up to the first order in $\delta Z$.  ${\cal H}$ is called Hessian
matrix.  One can easily integrate the above differential equation and
obtain the following solution:
\begin{equation}
	\delta Z(t)={\cal T}_\tau\exp\left[\int_{t_0}^t\d\tau
	\varLambda\,{\cal H}(\tau)\right]\,\delta Z(t_0)
	\equiv{\cal S}_\alpha(t-t_0)\,\delta Z(t_0),
\end{equation}
where ${\cal T}_\tau$ denotes that the exponential is defined by
time-ordered product.  ${\cal S}_\alpha$ is called stability matrix of
the trajectory $\alpha$, whose eigenvalues determine its stability.

Let $\varSigma$ denote a $(2f-2)$ dimensional hypersurface in the
phase space with fixed energy $E$.  It defines a time-discretized
mapping ${\cal M}:\varSigma \mapsto \varSigma$ with classical
trajectories, which is called the Poincar\'e map.  Periodic orbits
$\bar{Z}$ are defined as the fixed points of $\cal M$, namely, by
${\cal M}(\bar{Z})=\bar{Z}$.  The linear part $M_r$ (with respect to
$\delta Z$) of $\cal M$ about a periodic orbit $\bar{Z}_r$ is called
``monodromy matrix'' and describes the stability of the orbit:
\begin{equation}
	{\cal M}(\bar{Z}_r+\delta Z)
	=\bar{Z}_r + M_r\,\delta Z + {\cal O}(\delta Z^2).
\end{equation}
The monodromy matrix is a symplectic matrix satisfying
\begin{equation}
	\varLambda\,{M_r}^T\,\varLambda^{-1}={M_r}^{-1},
\label{eq:symplectic}
\end{equation}
and this property restricts its eigenvalues as follows.  Let $\lambda$
be one of the eigenvalues of $M_r$.  Relation (\ref{eq:symplectic})
guarantees that the reciprocal of $\lambda$ is another eigenvalue of
$M_r$.  Furthermore, $M_r$ is a real matrix so that the complex
conjugates of these eigenvalues are also eigenvalues.  Thus the
eigenvalues of the monodromy matrix generally appear in quartets
$(\e^{\pm\alpha\pm i\beta})$.  In the two-dimensional case, they
appear in a pair.

Let us now proceed to a discussion on bifurcations of stable periodic
orbits.  Consider a trajectory that emerges at $\vec{q}$ with energy
$E$ and returns to the initial point, $\vec{q}'=\vec{q}$.  This kind
of trajectory certainly exists for any $\vec{q}$.  The condition for
this trajectory to be periodic is that the initial and final momenta
coincide; namely, $\vec{p}'=\vec{p}$.  Using the Hamilton-Jacobi
equation, this condition can be rewritten as
\begin{equation}
	0=\vec{p}'-\vec{p}
	=\frac{\partial S(\vec{q}',\vec{q};E)}{\partial\vec{q}'}
	-\left(-\,\frac{\partial S(\vec{q}',\vec{q};E)}{
	\partial\vec{q}}\right)
	=\frac{\partial S(\vec{q})}{\partial\vec{q}},
\end{equation}
where $S(\vec{q})$ denotes the action integral along the closed path
under consideration.  Thus, the periodic orbit is the stationary point
of this action integral.  Let us expand $S(\vec{q})$ about the
periodic point $\bar{\vec{q}}$:
\begin{eqnarray}
	S(\bar{\vec{q}}+\delta\vec{q})&=&S(\bar{\vec{q}})
	+\frac12\,\delta\vec{q}^T \frac{\partial^2
	S(\vec{q},\vec{q})}{\partial\vec{q}\partial\vec{q}}\,
	\delta\vec{q}+\cdots
	\nonumber \\
	&\equiv& \bar{S}+\frac12\,\delta\vec{q}^T{\cal B}\,
	\delta\vec{q}+\cdots.
\end{eqnarray}
After some simple matrix rearrangements, one can express $\cal B$ in
terms of the quadrants of the monodromy matrix as
\begin{equation}
	{\cal B}=B-(I-A)C^{-1}(I-D), \qquad
	M \equiv \left(\begin{array}{c@{\quad}c}
	A & B \\ C & D
	\end{array}\right).
\end{equation}
Note that
\begin{equation}
	\det(\vec{1}-M)=-\det[C]\det[{\cal B}].
\end{equation}
The above relations provide us with a clear understandings of the
connection between eigenvalues of the monodromy matrix and the
bifurcation of periodic orbits.  Suppose that one of the eigenvalues
of $M$ becomes unity.  Then the curvature tensor $\cal B$ for the
action $S$ has a zero eigenvalue.  This means that the stationary
points of $S$ (periodic orbits) locally form a continuous set and a
bifurcation can occur hereafter; namely a (few) new stationary
point(s) can emerge.

\subsection{\label{SEC:TRFORMULA}Trace formula}

By means of the semiclassical theory, we can relate properties of the
quantum spectrum with those of the corresponding classical system.
For non-integrable Hamiltonian systems, the Gutzwiller's trace
formula \cite{Gutz:67,GutzText,Ozo:88} represents the quantum level
density $g(E)=\sum_i\delta(E-E_i)$ as a sum over classical periodic
orbits:
\begin{equation}
	g^{\rm(sc)}(E)=\bar{g}(E)+\sum_{r,n}
	A_{nr}(E)\cos\left(
	\frac{nS_r(E)}{\hbar}-\frac{\pi}{2}\mu_{nr}\right).
\label{eq:traceformula}
\end{equation}
$\bar{g}(E)$ is called Weyl term (or Thomas-Fermi approximation),
which is a monotonic function of energy.  The sum is taken over all
primitive periodic orbits $r$ and their multiple traversals.
$S_r=\oint_r\vec{p}\cdot d\vec{q}$ is the action integral along the
orbit $r$, and $\mu_r$ denotes the Maslov phase.  Several numerical
application of this formula to strongly chaotic systems have shown its
effectiveness.  As is well known, however, exact reproduction of a
quantum spectrum is not an easy task, because one has to treat huge
number of periodic orbits which show exponential proliferation as a
function of energy.  On the other hand, since our purpose is to
understand the gross structure of the spectrum we need only finite
number of periodic orbits which have rather short periods.  If one is
interested in the gross structure with energy resolution $\delta E$,
the change of the phase in Eq.~(\ref{eq:traceformula}) must be less
than $2\pi$ for the change of energy by $\delta E$.  Namely,
\begin{equation}
	nS_r(E+\delta E)-nS_r(E)
	\simeq n\frac{\partial S_r(E)}{\partial E}\delta E
	=nT_r \delta E \lesssim 2\pi\hbar,
\end{equation}
which leads to the relation
\begin{equation}
	nT_r \lesssim T_{\rm max}\equiv\frac{2\pi\hbar}{\delta E}.
\end{equation}
Thus, we need only short periodic orbits whose periods are less than
$T_{\rm max}$ defined above.  Although much efforts have been devoted
to reproduce individual eigenenergies by calculating millions of
periodic orbits, gross structures of the level spectra have rarely
been discussed in connection with periodic orbits for non-integrable
systems.
(For integrable systems, there are several works; see for instance,
Refs.~\CITE{NHM:90,BB:69,BT:76,MKS:78,Fr:90})

\subsection{\label{SEC:BIF}Bifurcations}

Let us next discuss the condition for the amplitude factor $A_{nr}$ in
the trace formula (\ref{eq:traceformula}) to take a large value.  In
the stationary phase approximation, the amplitude factor is expressed
for isolated orbits as
\begin{equation}
	A_{nr}=\frac{1}{\pi\hbar}
		\frac{T_r}{\sqrt{|\det(\vec{1}-{M_r}^n)|}}.
\end{equation}
For degenerate orbits with degeneracy 1, it is represented by
\begin{equation}
	A_{nr}=\frac{4\pi}{(2\pi\hbar)^{3/2}}
		\frac{B_r}{\sqrt{|\det(\vec{1}-{M_r}^n)|}}, \qquad
	B_r=\int_0^{T_r}\d t \left[\frac{
		\partial\varphi(t+T_r)}{\partial p_\varphi(t)}
		\right]^{-1/2},
\label{eq:amp-deg}
\end{equation}
where $\varphi$ and $p_\varphi$ denote an ignorable variable in the
Hamiltonian and its canonically conjugate momentum, respectively (see
Appendix in Ref.~\CITE{AM:94}).  From the difference in order of
$\hbar$, one sees that the contribution of degenerate orbits is more
important than that of isolated ones.

Another important factor, which plays an essential role in our
analysis below, is the stability factor $\det(\vec{1}-{M_r}^n)$.
Its value is independent of the point chosen on the periodic orbit.
As discussed above, eigenvalues of the monodromy matrix $M_r$ appear
in pairs $(+/-)(\e^\lambda, \e^{-\lambda})$, $\lambda$ being real or
pure imaginary, or in quartets $(\e^{\pm\alpha\pm i\beta})$.  One
should note that the periodic orbit generally appears in at least one
parameter family, so that the monodromy matrix always has two unit
eigenvalues.  Other pairs of unit eigenvalues correspond to the global
continuous symmetries which the Hamiltonian possesses but the orbit
itself does not.  These degrees of freedom are responsible for the
degeneracy and, as seen in Eq.~(\ref{eq:amp-deg}), can be separated
out from the definition of $M_r$ appearing in the stability factor.
In the case of three-dimensional systems with axial symmetry, the
monodromy matrix has (except for isolated non-degenerate orbits) four
unit eigenvalues, and the remaining two appear in a pair
$(+/-)(\e^\lambda, \e^{-\lambda})$.  Thus, $M_r$ may be reduced to a
(2$\times$2) matrix for most orbits.  $\lambda$ is purely imaginary
for stable orbits, and real for unstable ones.  The stability factor
for each case becomes
\begin{displaymath}
	\det(\vec{1}-M_r)=2-\Tr M_r
\end{displaymath}
with
\begin{equation}
	\Tr M_r=\left\{
	\begin{array}{l@{\qquad}l}
	2\cos(\beta) & \mbox{stable},\quad \lambda=i\beta \\
	\pm2\cosh(\alpha)) & \mbox{unstable},\quad \lambda=\alpha
	\end{array}
	\right.
\label{eq:tracem}
\end{equation}
Thus, the stability of orbit is determined by the value of $\Tr M$.

If a parameter in the Hamiltonian is continuously varied, the periodic
orbits change their shapes and the values of $\lambda$ also change
continuously.  It may occur that the $\beta$ for a certain stable
orbit becomes a fraction of $2\pi$, namely, $\beta=2\pi m/n$ with $n$
and $m$ being relatively prime integers.  At this point $\Tr {M_r}^n$
becomes 2, and the amplitude factor $A_{nr}$ suffers divergence.  This
singularity corresponds to the period $n$-upling bifurcation of the
orbit $r$.  Near the bifurcation point, the stationary phase
approximation breaks down.  It is then necessary to take into account
higher-order fluctuations about the classical orbit to extract a
finite value of $A_{nr}$ \cite{OH:87}.
Although we leave this task as an challenging future subject, we
expect that the amplitude factor takes large value in the bifurcation
region.  It will result in a large-amplitude oscillation in the level
density, leading to an enhancement of the shell effect.

\section{\label{SEC:MODEL}The model and its scaling properties}

\subsection{The model}

We adopt a model Hamiltonian consisting of an
axially deformed harmonic oscillator and a reflection asymmetric
octupole deformed potential:
\begin{equation}
	H=\frac{~p^2}{2M}+\sum_i\frac{M\omega_i^2x_i^2}{2}
	-\lambda_{30}M\omega_0^2\left[r^2Y_{30}(\varOmega)\right]''.
\label{eq:hamiltonian}
\end{equation}
Here, the double primes denote that the variables in square brackets
are defined in terms of the doubly-stretched coordinates
$x_i''=(\omega_i/\omega_0)x_i$, where $\omega_0\equiv(\omega_x \omega_y
\omega_z)^{1/3}$ being determined so that the volume
conservation condition is satisfied \cite{AM:93}.  For simplicity, we
define dimensionless variables as follows:
\begin{equation}
	\left\{
	\begin{array}{lll}
	p_i &\longrightarrow &\sqrt{M\hbar\omega_0}\,p_i  \\
	q_i &\longrightarrow &\sqrt{\hbar/M\omega_0}\,q_i \\
	H   &\longrightarrow &\hbar\omega_0\,H
	\end{array}
	\right.
\end{equation}
Then, the Hamiltonian is written as
\begin{equation}
	H=\frac12\,p^2+\frac12\,[r^2(1-2\lambda_{30}Y_{30}(\varOmega))]''.
\label{eq:shamiltonian}
\end{equation}

Since the radial dependence of the octupole potential is quadratic,
the Hamiltonian obeys the following scaling rule:
\begin{equation}
	H(\alpha\vec{p}, \alpha\vec{q})
	=\alpha^2 H(\vec{p},\vec{q}).
\label{eq:scaling}
\end{equation}
Thanks to this property, if one solves the Hamilton equations of
motion at a certain energy $E_0$, the solution at any energy $E$ is
obtained by just scale transforming the solution at $E_0$ as
$Z(t;E)=\sqrt{E/E_0}\,Z(t;E_0)$, $Z$ denoting a phase space vector
$(\vec{p},\vec{q})$.

In the cylindrical coordinates $(\rho, z, \varphi)$, the above
Hamiltonian (\ref{eq:shamiltonian}) is written as
\begin{displaymath}
	H=\frac12(p_\rho^2+p_z^2)+V(\rho, z; p_\varphi),
\end{displaymath}
\begin{equation}
	V(\rho, z; p_\varphi)=\frac{p_\varphi^2}{2\rho^2}
	+\left[\frac{\rho^2+z^2}{2}-\lambda_{30}\sqrt{\frac{7}{16\pi}}
	\frac{2z^3-3z\rho^2}{\sqrt{\rho^2+z^2}}\right]''.
\label{eq:cylindrical}
\end{equation}
Thus, we can treat the system as a two-dimensional one with fixed
angular momentum $p_\varphi$.  This reduction enables us to make use
of the Poincar\'e surface of section in order to survey the classical
phase space profile.  Note that $\chi\equiv p_\varphi/E$ is a
scaling-invariant parameter and, therefore, classical properties are
the same for the same $\chi$.

In the following sections, we shall investigate how the properties of
the quantum spectrum for the Hamiltonian (\ref{eq:hamiltonian})
changes as the two deformation parameters, $\delta_{\rm
osc}=(\omega_\perp-\omega_z)/\bar{\omega}$ and $\lambda_{30}$, are
varied.  We shall then discuss the physical origins of these changes
by means of the periodic orbit theory reviewed in section
\ref{SEC:FORM}.

\subsection{\label{SEC:FOURIER}
Fourier transformation of quantum level density}

As we will see in the following numerical analyses, the Hamiltonian
above becomes chaotic with increasing octupole deformation parameter
$\lambda_{30}$.  But considerable parts of the phase space remain
regular and the system is considered as a so-called `mixed system'.
The trace formula based on the stationary phase approximation (SPA)
does not work well in such situations.  The amplitude factors suffer
divergences at the bifurcation points of stable periodic orbits
because of the breakdown of SPA.  Consequently, we cannot directly use
the conventional semiclassical expression to analyze the quantum
spectrum.  Fortunately, we can avoid the above difficulty by using the
Fourier transformation technique for the quantum level density.
Suppose that the level density is characterized by the classical
periodic orbits and is expressed as
\begin{equation}
	g_{\rm osc}^{\rm (sc)}(E)
	=\sum_{n=1}^\infty \sum_r A_{nr}(E)\,
	\cos\left(\frac{nS_r(E)}{\hbar}-\frac{\pi}{2}\mu_{nr}\right),
\label{eq:scgosc}
\end{equation}
without specifying the concrete expression of the amplitude factor
$A_{nr}(E)$ which may be obtained by going beyond the SPA.  Since our
model obeys the scaling rule (\ref{eq:scaling}), energy dependence of
the classical variables entering Eq.~(\ref{eq:scgosc}) is factored out
as follows:
\begin{equation}
	\begin{array}{l}
	S_r(E)=E\cdot T_r , \\
	A_{nr}(E)=E^{d_r/2}A_{nr}^{(0)} , \\
	\end{array}
\label{eq:scale}
\end{equation}
where $d_r$ is the effective degeneracy of the orbit, and equal to 1
for most orbits due to the axial symmetry.  For isolated orbits,
$d_r=0$ and we expect that their contributions to the level density
may be small.  The degeneracies are integers in the classical
dynamics, but in the quantum mechanics this restriction is relaxed and
$d_r$ changes continuously in the bifurcation regions
\CITE{OH:87,Ozo:88}.  Using the above relations, Eq.~(\ref{eq:scale})
is expressed as
\begin{equation}
	g^{\rm(sc)}(E)=\bar{g}(E)+\sum_{r,n}E^{d_r/2}A_{nr}^{(0)}
	\cos\left(\frac{nET_r}{\hbar}-\frac{\pi}{2}\mu_{nr}\right).
\label{eq:ldsemicl}
\end{equation}

Now let us consider the Fourier transformation of the level
density, defined by
\begin{equation}
	F(s)=\frac{1}{2\pi\hbar}
	\int\d E\,\e^{isE/\hbar} E^{-d/2} g(E)
	\cdot\exp\left[-\frac12\left(
	\frac{E}{E_{\rm max}}\right)^2\right].
\label{eq:fourier}
\end{equation}
Here the Gaussian damping factor is used for energy cut-off, and we
shall put $d=1$ in order to cancel the energy dependence of the
amplitude factors for most orbits.  Inserting the quantum level
density $g(E)=\sum_i\delta(E-E_i)$ and the semiclassical one
(\ref{eq:ldsemicl}) into Eq.~(\ref{eq:fourier}), we obtain quantum
mechanical and semiclassical expressions for $F(s)$:
\begin{equation}
	F^{\rm(qm)}(s)=\frac{1}{2\pi\hbar}
	\sum_i \frac{1}{\sqrt{E_i}} \e^{isE_i/\hbar}
	\cdot\exp\left[-\frac12\left(
	\frac{E_i}{E_{\rm max}}\right)^2\right],
\label{eq:qmfourier}
\end{equation}
\begin{equation}
	F^{\rm(sc)}(s)\simeq\bar{F}(s)
	+\sum_{r,n}A_{nr}^{(0)}
	\cdot\frac{1}{\sqrt{2\pi}\varDelta s}
	\exp\left[-\frac12 \left(
	\frac{s-nT_r}{\varDelta s}\right)^2\right],
\label{eq:scfourier}
\end{equation}
respectively, where $\varDelta s=\hbar/E_{\rm max}$.  $F^{\rm(qm)}$ is
calculated from the single-particle spectrum obtained by diagonalizing
the Hamiltonian with deformed oscillator basis.  The result is
compared with the semiclassical expression $F^{\rm(sc)}$.  In
Eq.~(\ref{eq:scfourier}), $\bar{F}(s)$ corresponds to the Weyl term
which is regarded as a contribution from orbit of zero-length, and it
has peak at $s=0$.  The remaining part has a functional form
exhibiting successive peaks at the periods of classical periodic
orbits and their heights are proportional to the amplitude factors of
the corresponding orbits.  By comparing the calculated
$F^{\rm(qm)}(s)$ with $F^{\rm(sc)}(s)$, we can thus extract
information about periodic orbits from the quantum spectrum.

\section{\label{SEC:ESH}Shell structure energy calculation}

A useful quantitative measure of shell structure is the shell
structure energy which is defined as the fluctuation part of the sum
of single-particle energies, i.e.,
\begin{equation}
	{\cal E}_{\rm sh}(N)=\sum_{k=1}^N E_k - \tilde{\cal E}(N),
\end{equation}
where $E_k$'s represent eigenvalues of the single-particle
Hamiltonian.  The second term $\tilde{\cal E}(N)$ is obtained by
smoothing the first term by means of the Strutinsky method; it is a
smooth function of particle number $N$ and of other potential
parameters.  The shell structure energy takes a large negative value
when the single-particle level density at the Fermi surface is low.

We have carried out a systematic calculation of the shell structure
energy for the Hamiltonian (\ref{eq:hamiltonian}) as a function of the
deformation parameters $\delta_{\rm osc}$ (quadrupole deformation),
$\lambda_{30}$ (octupole deformation) and of the particle number $N$.
\ifepsf
\begin{figure}[tb]
\else
\begin{figure}[e]
\fi
\epsfxsize=\textwidth
\epsffile{vmin.ps}
\caption{\label{fig:vmin}
Local minima of shell structure energy in the two-dimensional
deformation parameter space ($\delta_{\rm osc}, \lambda_{30}$).  Size
of each disc represents the absolute value of the shell structure
energy normalized as ${\cal E}_{\rm sh}/N^{1/3}$.  Plotted are for
even $N$ in the range $16\leq N\leq 160$.}
\end{figure}
Figure~\ref{fig:vmin} shows a map of the local minima of shell
structure energies in the two-dimensional deformation parameter space
calculated for particle numbers in the range $16\leq N\leq 160$.  The
centers of discs show the loci of local minima (corresponding to
different values of $N$) and the sizes of the discs represent the
absolute values of shell structure energies.  As the order of
magnitude of the shell structure energy is roughly proportional to
$N^{1/3}$ in the harmonic oscillator case \cite{BM:72}, we normalize
them by multiplying $N^{-1/3}$ in order to compare system with
different values of $N$.

For $\lambda_{30}=0$, it is well known that strong shell structure
exists at $\delta_{\rm osc}=0$ (spherical), 3/5 (prolate
superdeformed), 6/7 (prolate hyperdeformed), $-3/4$ (oblate
superdeformed) and so on.  The distribution of discs is actually dense
around such points.  We find in Fig.~\ref{fig:vmin} that, in addition
to such known cases, prominent shell structure emerges also at finite
values of octupole deformation.  Their shell-structure energies are
comparable, in magnitude, to (sometimes larger than) those for the
purely quadrupole shapes.  The most remarkable region is that of
$\lambda_{30}=0.3\sim0.4$ and $\delta_{\rm osc}\simeq0.1$.  Using the
semiclassical theory, let us analyze in the following sections the
mechanism which creates the above new shell structures for the
combination of the quadrupole and octupole deformations.

\section{\label{SEC:SEMICL}Semiclassical analyses}

In this section, we investigate how the structure of quantum spectrum
changes as the octupole deformation parameter $\lambda_{30}$ is
increased, fixing quadrupole deformation parameter $\delta_{\rm osc}$
at several positive values (between the spherical and the prolate
superdeformed shapes).  Figure~\ref{fig:nils-p} shows the
single-particle spectra as functions of $\lambda_{30}$ for the
Hamiltonian (\ref{eq:hamiltonian}) with $\delta_{\rm osc}=0.1$, 0.3
and 0.5,
\begin{figure}[p]
\begin{minipage}{.495\textwidth}
\epsfxsize=\textwidth\epsffile{nils-D10p.ps}
\end{minipage}
\hfill
\begin{minipage}{.495\textwidth}
\epsfxsize=\textwidth\epsffile{nils-D30p.ps}
\end{minipage}
\par\vspace{16pt}
\begin{minipage}{.495\textwidth}
\epsfxsize=\textwidth\epsffile{nils-D50p.ps}
\end{minipage}
\hfill
\begin{minipage}[t]{.47\textwidth}
\caption{\label{fig:nils-p}
Single-particle spectrum of the Hamiltonian
(\protect\ref{eq:hamiltonian}) with deformation parameter $\delta_{\rm
osc}=0.1$, 0.3 and 0.5 as functions of the octupole parameter
$\lambda_{30}$.  Dashed and solid curves represent the levels whose $K$
quantum numbers are zero and nonzero, and the latter degenerate in two
due to the time-reversal symmetry.}
\end{minipage}
\end{figure}
The corresponding axis ratios $\omega_\perp/\omega_z$ are
%
%
31/28, 11/8 and 7/4, respectively.  They are not in simple ratios so
that there is no prominent shell structure at $\lambda_{30}=0$.
However, new shell structures emerge at finite values of
$\lambda_{30}$.  To see features of these spectra, we show in
Fig.~\ref{fig:ftlmap-p} the Fourier transforms of the level density
defined by (\ref{eq:qmfourier}).
\begin{figure}[p]
\leftline{(a) $\delta_{\rm osc}=0.1$}
\vspace{-\baselineskip}
\epsfxsize=\textwidth
\epsffile{ftl-D10p.ps}\par
\leftline{(b) $\delta_{\rm osc}=0.3$}
\vspace{-\baselineskip}
\epsfxsize=\textwidth
\epsffile{ftl-D30p.ps}
\centerline{\small Fig.~\protect\ref{fig:ftlmap-p}, to continue.}
\end{figure}
\begin{figure}[tb]
\leftline{(c) $\delta_{\rm osc}=0.5$}
\vspace{-\baselineskip}
\epsfxsize=\textwidth
\epsffile{ftl-D50p.ps}
\caption{\label{fig:ftlmap-p}
Fourier transforms $|F(s)|$ of the level densities for the Hamiltonian
(\protect\ref{eq:hamiltonian}) with $\delta_{\rm osc}=0.1$, 0.3 and
0.5, plotted as functions of the action $s$ and the octupole parameter
$\lambda_{30}$.}
\end{figure}
As discussed in section \ref{SEC:FOURIER}, one see prominent Fourier
peaks at the periods of classical periodic orbits.  It means that the
fluctuation of the spectrum is characterized by the periodic orbits,
demonstrating a beautiful applicability to our model of the
semiclassical method in section \ref{SEC:TRFORMULA}.  For elucidating
the features of shell structure, it is essential to understand the
behavior of these Fourier peaks with respect to the deformation
parameters.  As discussed in the previous sections, short periodic
orbits play important roles for the formation of gross structure of
quantum spectrum.  There are various periodic orbits of various
topologies in each part of the deformation parameter space and the
same type of orbits change their characters as the parameters change.
This fact is clearly seen in the Fourier transforms where peaks
corresponding to certain orbits change their heights.  The heights of
the peaks represent nothing but the strengths of shell structures.
Let us discuss in the followings what kind of periodic orbits exist
and how they determine the features of quantum spectra in several
regions on the $(\delta_{\rm osc},\lambda_{30})$ plane.

\subsection*{\underline{Case of $\delta_{\rm osc}=0.1$}}

Let us first take up the case of $\delta_{\rm osc}=0.1$, where we
obtain especially strong shell structures at finite $\lambda_{30}$
values in the shell structure energy calculation (see
Fig.~\ref{fig:vmin}), and let us discuss which orbits are responsible
for these shell structures.  Figure~\ref{fig:po-zpo} shows, for
several values of $\lambda_{30}$, some planar periodic orbits in the
plane including the symmetry axis.
We use the Monodromy Method developed by Baranger et al. \cite{BDM:88}
to calculate periodic orbits and their monodromy matrices.
At $\lambda_{30}=0$, the most important orbit family is the
ellipse-shaped one in the $(x,y)$ plane.  The next orbit family is
31:28 Lissageous.  They are very long orbits and unimportant for the
gross shell structure.  Adding the octupole deformation, new types of
orbits are born.  Orbit PR appears at $\lambda_{30}=0.12$ by the
isochronous bifurcation of the orbit PA, and PA becomes unstable after
this bifurcation.  At $\lambda_{30}=0.24$, the orbit PA becomes stable
again and a new orbit PM appears.

\begin{figure}[p]
\epsfxsize=.75\textwidth
\centerline{\epsffile{po-D10p.ps}}
\caption{\label{fig:po-zpo}
Some short planar periodic orbits of the Hamiltonian
(\protect\ref{eq:hamiltonian}) with $\delta_{\rm osc}=0.1$.}

\vspace{20pt}
\epsfxsize=\textwidth
\centerline{\epsffile{pmap-D10p.ps}}
\caption{\label{fig:pmap-zpo}
Poincar\'e surface of section $(z,p_z)$ for the Hamiltonian
(\protect\ref{eq:cylindrical}) with $p_\varphi=0$.  The surface is
defined by $p_\rho=0$ and $\dot{p}_\rho<0$.}

\vspace{20pt}
\epsfxsize=.6\textwidth
\centerline{\epsffile{trm-D10p.ps}}
\begin{center}
\begin{minipage}{.7\textwidth}
\caption{\label{fig:trm-zpo}
Traces of the Monodromy matrices at $\delta_{\rm osc}=0.1$ for some
short periodic orbits plotted as functions of the octupole deformation
parameter $\lambda_{30}$.}
\end{minipage}
\end{center}
\end{figure}

To see these bifurcations, the Poincar\'e map is a very convenient
implement.  As our Hamiltonian has axial symmetry, we can treat it as
a two-dimensional system with fixed angular momentum $p_\varphi$.
Figure~\ref{fig:pmap-zpo} is the Poincar\'e map $(z,p_z)$ for the
`projected' Hamiltonian (\ref{eq:cylindrical}) with $p_\varphi=0$.
Here the surface is defined by $p_\rho=0$ and $\dot{p}_\rho<0$.  In
the top panel, the center of the main torus corresponds to the orbit
PA.  In the middle panel, the orbit PA becomes an unstable saddle and
a new pair of island is created.  These islands correspond to the
orbit PR.  In the bottom panel the orbit PA comes back to a stable one
and creates a new pair of saddles confronting horizontally.  These
saddles correspond to the orbit PM.

As is well known, stabilities of periodic orbits are determined by the
monodromy matrices.  Figure~\ref{fig:trm-zpo} shows the traces of the
two-dimensional monodromy matrices for orbits associated with the
above bifurcations.  The bifurcation occurs at $\Tr M=2$ where the
monodromy matrix has the unit eigenvalues $(1,1)$.  Period $n$-upling
bifurcations occur at $\Tr M^n=2$.  Higher order (large $n$)
resonances occur at every points of the torus and new orbits bifurcate
from them, but they are of rather long periods and are related with
more detailed structure of the spectrum.  In the Fourier transform
Fig.~\ref{fig:ftlmap-p}(a), the peak corresponding to the orbit PA
with period $T\simeq2\pi/\omega_\perp$ strongly enhances at
$\lambda_{30}\simeq0.2$.  It mainly characterizes the shell structure
seen in the spectrum Fig.~\ref{fig:nils-p}(a) at
$\lambda_{30}=0.2\sim0.3$.  This enhancement is related with the above
bifurcations of PA at $\lambda_{30}=0.128$ and 0.227.

\subsection*{\underline{Case of $\delta_{\rm osc}=0.3$}}

Next let us discuss the case of $\delta_{\rm osc}=0.3$.  Some
important classical orbits are drawn in Fig.~\ref{fig:po-zpt}.
\begin{figure}
\epsfxsize=.9\textwidth
\centerline{\epsffile{po-D30p.ps}}
\caption{\label{fig:po-zpt}
Some short planar periodic orbits for the Hamiltonian
(\protect\ref{eq:hamiltonian}) with $\delta_{\rm osc}=0.3$.}
\end{figure}
Orbit PO is born out of the period-doubling bifurcation of the orbit
IL at $\lambda_{30}=0.215$.  Orbits PP and PQ are created by the
saddle-node bifurcation (pair creation of stable and unstable orbits
from nothing) at $\lambda_{30}=0.221$.  Orbit PR emerges from the
isochronous bifurcation of the orbit PA at $\lambda_{30}=0.338$.  In
the Fourier transform Fig.~\ref{fig:ftlmap-p}(b), the enhancement of
the peak with $s\simeq 1$ is related with the isochronous bifurcation.
The orbits associated with the period-doubling and saddle-node
bifurcations mentioned above are of similar periods and contribute to
the same peak with $s\simeq2.75$ in the Fourier transform.  It also
shows strong enhancement in the bifurcation region.

\subsection*{\underline{Case of $\delta_{\rm osc}=0.5$}}

Lastly, let us consider the case of $\delta_{\rm osc}=0.5$.  This
quadrupole deformation is almost equivalent to the case of axis ratio
$\sqrt{3}:1$ treated in Ref.~\CITE{Ari:94}.  The feature of shell is
rather weak at $\lambda_{30}=0$.  However, one may notice in
Fig.~\ref{fig:nils-p} that a remarkable shell structure emerges at
$\lambda_{30}\simeq0.3$.  In the Fourier transform
Fig.~\ref{fig:ftlmap-p}(c), a significant enhancement of the peak with
$s\simeq1.75$ is observed.  It seems that this is the most typical
example in our model, which clearly exhibits a new shell structure
emerging at finite octupole deformation.

Some short periodic orbits are illustrated in Fig.~\ref{fig:po-zpf}.
\begin{figure}
\epsfxsize=.9\textwidth
\centerline{\epsffile{po-D50p.ps}}
\caption{\label{fig:po-zpf}
Some short planar periodic orbits for the Hamiltonian
(\protect\ref{eq:hamiltonian}) with $\delta_{\rm osc}=0.5$.}

\vspace{20pt}
\epsfxsize=\textwidth
\epsffile{pmap-D50p.ps}
\caption{\label{fig:pmap-zpf}
Poincar\'e surfaces of section for $\delta_{\rm osc}=0.5$ in the
bifurcation region of short periodic orbits.  The surface
$(\rho, p_\rho)$ is defined by $z=0$ and $p_z>0$.}
\end{figure}
At $\lambda_{30}=0$, the ellipse-shaped family in the $(x,y)$ plane
characterize the structure of the spectrum.  Orbit PB is born out of
the isochronous bifurcation of the orbit IL at $\lambda_{30}=0.276$,
orbits PC and PD are created by the saddle-node bifurcation at
$\lambda_{30}=0.274$.  The Poincar\'e maps and the traces of the
monodromy matrices are displayed in Figs.~\ref{fig:pmap-zpf} and
\ref{fig:trm-zpf}, respectively.
\begin{figure}
\epsfxsize=.6\textwidth
\centerline{\epsffile{trm-D50p.ps}}
\caption{\label{fig:trm-zpf}
Same as Fig.~\protect\ref{fig:trm-zpo} but $\delta_{\rm osc}=0.5$.}
\end{figure}
The strong enhancement of the peak with $s\simeq1.85$ at
$\lambda_{30}\simeq0.4$ are related with the bifurcations mentioned
above.  Orbits associated with these bifurcations are of similar
periods and thus contribute to the same peak in the Fourier transform
Fig~\ref{fig:ftlmap-p}(c).  These bifurcations may be regarded as the
mechanism which creates the prominent shell structure at finite
octupole deformation.

Orbits PE and PF are born associated with the period-tripling
bifurcation of the orbit PA at $\lambda_{30}=0.376$.  A quantum
signature of this bifurcation is also seen in the Fourier peak with
$s\simeq3$.  Thus, we can conclude that the enhancement of Fourier
peak in the bifurcation region is a general phenomenon.

\section{\label{SEC:SYS}%
Bifurcation lines in the two-dimensional parameter space}

In the preceding section, we presented several examples where
bifurcations of short periodic orbits play important roles in forming
shell structures.  In this section, let us discuss roles of the
bifurcations varying two deformation parameters systematically.  In
the $n$-dimensional parameter space, bifurcation points generally form
$(n-1)$-dimensional manifolds; they are 1-dimensional curves in the
present case.  We have evaluated these bifurcation lines for some
short periodic orbits.  The result of calculation is presented in
Figure~\ref{fig:biflines}.  These bifurcation lines always emerge at
the points where the ratios $\omega_\perp/\omega_z$ are rational and
$\lambda_{30}=0$.  Note that $\delta_{\rm osc}$=0, 3/5, 6/7 and $-3/4$
correspond to the spherical, prolate superdeformed, prolate
hyperdeformed and oblate superdeformed shapes, respectively.
The orbit PA causes isochronous bifurcations along the lines {\tt i}
and {\tt ii} (in direction from the right to the left).  As discussed
in the case $\delta_{\rm osc}=0.1$, new orbits PM and PR are born
after the bifurcations {\tt i} and {\tt ii}, respectively.  On the
other hand, as discussed in the case $\delta_{\rm osc}=0.5$, the lines
{\tt iii}$\sim${\tt vi} are related with the orbits PB, PC, PD, IL and
2PA (double traversals of the orbit PA).  The orbit IL causes
isochronous bifurcation (from the left to the right) and produces the
orbit PB along the line {\tt iii}, while the orbit PC (unstable) and
PD (stable) are born out of the saddle-node bifurcation along the line
{\tt iv} in the region $\delta_{\rm osc}\lesssim0.5$.  As $\delta_{\rm
osc}$ increases, orbits PB and PC exchange their stabilities with each
other along the line {\tt v} and annihilate into 2PA along the line
{\tt vi}.  Along the line {\tt vii}, the orbit PA causes
period-tripling bifurcation (from the right to the left) and produces
orbits PE and PF.

\begin{figure}[tb]
\epsfxsize=\textwidth
\epsffile{biflines.ps}
\caption{\label{fig:biflines}
Bifurcation lines for some short periodic orbits in the
two-dimensional deformation parameter space ($\delta_{\rm osc},
\lambda_{30}$).}
\end{figure}

According to the discussions in the preceding sections, we expect
strong shell effects along these bifurcation lines.  Prominent shell
structures are known to exist for the spherical and superdeformed
shapes (the end points of the bifurcation lines) where periodic orbit
conditions for the harmonic oscillator potential are satisfied.
Conditions for the emergence of pronounced shell structure in
non-integrable systems are not well known, however.  As is indicated
in Fig.~\ref{fig:vmin}, shell structure energies at equilibrium shapes
with finite $\lambda_{30}$ are comparable in magnitude to those of
spherical and superdeformed shapes.  The most remarkable one is the
region with $\delta_{\rm osc}\simeq0.1$ and $\lambda_{30}=0.3\sim0.4$.
The bifurcation map of Fig.~\ref{fig:biflines} suggests that the
strong shell effect in this region may be connected with the
bifurcation of orbit PA (the lines {\tt i} and {\tt ii}).  In fact, we
saw in the preceding section that the Fourier peak corresponding to
the orbit PA is strongly enhanced by the bifurcation phenomena.  It is
worth emphasizing that such a strong shell effect can arise associated
with classical orbit bifurcations in non-integrable systems.

One should also note that most of the large discs in
Fig.~\ref{fig:vmin} at finite $\lambda_{30}$ locate in the prolate
side ($\delta_{\rm osc}>0$).  This is related with the property of the
shortest periodic orbit: For prolate shapes, the shortest orbit is of
the type-PA.  It is degenerate and stable, so its contribution to the
level density is important.  On the other hand, the shortest orbit for
oblate shapes is of the type-IL, which is isolated and, accordingly,
its contribution to the level density is rather small.  The orbit of
type-PA, the second shortest orbit, is unstable against the octupole
deformation, and less important in comparison with the prolate case.
A more detailed discussion on the difference in stability between the
oblate and prolate superdeformed shapes against octupole deformations
will be given in the next section.  The same problem was discussed
also in \CITE{HNR:94a,HNR:94b} from somewhat different point of view.

\section{\label{SEC:SD}Octupole deformation superposed on the prolate
and oblate superdeformations}

Let us discuss the origin of the difference in octupole stability
between the prolate and oblate superdeformed states.  In
Ref.~\CITE{AM:93,AM:94} we have shown that the supershell effect in
the prolate superdeformed states increases with increasing octupole
deformation.  As an underlying mechanism of that enhancement, we
emphasized the importance of stability properties of two kinds of
periodic orbit family and of their interference effect.  The oblate
case is similar to the prolate case in that there are two kinds of
periodic orbit family whose periods are in the ratio 2:1.  But the
structure of quantum spectrum is quite different with each other.
\begin{figure}[bt]
\begin{minipage}{.495\textwidth}
\epsfxsize=\textwidth
\epsffile{nils-D60p.ps}
\centerline{\protect\footnotesize\sf (a)}
\end{minipage}
\hfill
\begin{minipage}{.495\textwidth}
\epsfxsize=\textwidth
\epsffile{nils-D75o.ps}
\centerline{\protect\footnotesize\sf (b)}
\end{minipage}
\caption{\label{fig:nils-sd}
Single-particle spectra for (a) prolate and (b) oblate superdeformed
oscillators plotted as functions of the octupole deformation parameter
$\lambda_{30}$.}
\end{figure}
In Fig.~\ref{fig:nils-sd} are compared the single-particle spectra for
the prolate $(\omega_\perp/\omega_z=2)$ and oblate
$(\omega_\perp/\omega_z=1/2)$ superdeformed oscillators as functions
of the octupole deformation parameter $\lambda_{30}$.  The way the
degeneracy be solved is different between the two.  The octupole
operator $Y_{30}$ has matrix elements between states in the same major
shell in the oblate case and, therefore, it affects the spectrum in
the first order perturbation, while it affects only in the second
order in the prolate case.  Let us discuss below how this difference
be explained in terms of the classical dynamics point of view.  For
this purpose, representative periodic orbits with short periods are
displayed in Fig.~\ref{fig:orbit-sd} for several octupole deformation
parameters $\lambda_{30}$.

We first compare the features of the two spectra without the octupole
term.  Figure \ref{fig:old-sd} shows the oscillating level density
smoothed to an energy width $\delta E=\omega_{\rm sh}/2$ ($\omega_{\rm
sh}$ being the energy spacing between adjacent major shells).
\begin{figure}[tb]
\begin{center}
\epsfxsize=.6\textwidth
\centerline{\epsffile{old-SD.ps}}
\begin{minipage}{.8\textwidth}
\caption{\label{fig:old-sd}
Oscillating level density of prolate and oblate superdeformed
oscillators, which are smoothed to an energy width
$\delta E=\hbar\omega_{\rm sh}/2$.}
\end{minipage}
\end{center}
\end{figure}
A characteristic feature is that the prolate superdeformed spectrum
has an undulating pattern (supershell structure) while oblate one does
not.  This is due to the difference in degeneracies of contributing
periodic orbits.
\begin{figure}[p]
\epsfxsize=.7\textwidth
\centerline{\epsffile{po-D60p.ps}}
\vspace{6mm}
\epsfxsize=.8\textwidth
\centerline{\epsffile{po-D75o.ps}}
\vspace{6mm}
\caption{\label{fig:orbit-sd}
Some short periodic orbits in the prolate (upper panel) and the oblate
(lower panel) superdeformed potentials with octupole deformations.}
\end{figure}
As discussed in Ref.~\CITE{AM:94}, two orbit families (corresponding
to orbits (PB, PC) and orbit PA in the upper panel of
Fig.~\ref{fig:orbit-sd}, respectively) in the prolate superdeformed
states have degeneracies 4 and 2.  In the oblate case, the orbit
family with period $2\pi/\omega_\perp$ (corresponding to orbits (PA,
PL) in the lower panel of Fig.~\ref{fig:orbit-sd} has the maximal
degeneracy 4, but the shortest orbit (the linear orbit IL along the
$z$-axis) is isolated and has degeneracy 1.  Thus the interference
effect between these two families is so small that one cannot see the
supershell effect in the spectrum.

Figure~\ref{fig:ftlmap-sd} compare the Fourier transforms of the level
density for the prolate and oblate cases.
\begin{figure}[p]
\epsfxsize=\textwidth
\epsffile{ftl-D60p.ps}
\epsfxsize=\textwidth
\epsffile{ftl-D75o.ps}
\caption{\label{fig:ftlmap-sd}
Fourier transforms of the quantum level densities for the prolate
(upper panel) and the oblate (lower panel) superdeformed oscillators
with octupole deformations.}
\end{figure}
One can hardly see the component at $s=1$ in the oblate case and the
oscillating pattern of the spectrum is determined by the $s=2$
component almost exclusively.  Comparing the two figures, one notice
that the reduction rate of the peak-height due to the octupole
deformation is much greater in the oblate case.  This rapid decline
clearly corresponds to the rapid disappearance of the shell effect in
the oblate case.

The main reason for reduction of the shell effect with increasing
octupole deformation is two-fold: The first is the reduction of
degeneracy of the periodic orbit families, and the second is the
change of stabilities.  As the degeneracies are the same for the major
orbit families in both cases (orbits PB, PC, $\cdots$ in the prolate
case and orbits PA, PL, $\cdots$ in the oblate case), we expect that
the differences are associated mainly with the stability properties.
In Fig.~\ref{fig:stb-sd} we show the stability factors
$X\equiv\sqrt{|\det(1-M_r)|}$ calculated as function of
$\lambda_{30}$.
\begin{figure}[tb]
\epsfxsize=.9\textwidth
\centerline{\epsffile{stb-SD.ps}}
\caption{\label{fig:stb-sd}
Stability factors $X=\protect\sqrt{|\det(1-M_r)|}$ of the short
periodic orbits for the prolate (left) and the oblate (right)
superdeformed potentials plotted as functions of the octupole
deformation parameter $\lambda_{30}$.  The sign of $X$ is that of
$\det(M_r-1)=\Tr M_r-2$.  Namely, $-2<X\leq0$ for stable orbits and
otherwise for unstable ones.}
\end{figure}
The stability factors for orbits in the oblate potential depend
linearly on $\lambda_{30}$ for $\lambda_{30}\simeq0$ while they depend
quadratically in the prolate case.  Consequently the amplitude factors
reduce much faster in the former case.  This seem to be the main cause
of the rapid reduction of shells with octupole deformation in the
oblate potential.

\section{\label{SEC:SUMMARY}Summary and conclusion}

We have analyzed the gross structure of single-particle spectra in
reflection-asymmetric deformed oscillator potentials using the
semiclassical method.  Our model is non-integrable and is regarded as
a mixed system where regular and chaotic dynamics coexist.  The
periodic orbit theory, which is well established for regular and
strongly chaotic limits, seems to be also applicable to such a
situation.  Fourier transforms of the quantum level density reveal
almost perfect correspondences with classical periodic orbits.
Importance of classical orbit bifurcations has been demonstrated in
our model.  Strong shell effects arise also for rather chaotic
regions, and their strengths are comparable in magnitude to those of
regular regions.  We obtain an interesting result which indicates that
classical bifurcations may be responsible for the emergence of shell
structure in the mixed system.  Applications of the semiclassical
theory of shell structure to more realistic mean-field potential
models and identifications of classical orbits which play decisive
roles in determining exotic shapes of nuclei or micro-clusters remain
as exciting future subjects.

\section*{Acknowledgements}

The authors thank Dr. M.~Matsuo (Yukawa Institute for Theoretical
Physics) and Dr. H.~Aiba (Koka Women's College) for fruitful
discussions.

%
%

\end{document}